\documentclass[%
 reprint,
 amsmath,amssymb,
 aps,
]{revtex4-2}

\usepackage{graphicx}
\usepackage{dcolumn}
\usepackage{bm}
\usepackage{hyperref}
\usepackage{xurl}

\usepackage{physics}

\begin{document}

\preprint{}

\title{Multi-Species Cohesion: Humans, machinery, AI and beyond}

\author{Frank Yingjie Huo}
\author{Pedro D. Manrique}
\author{Neil F. Johnson}
\affiliation{%
 Physics Department, George Washington University, Washington, DC 20052, U.S.A.
}%

\date{\today}

\begin{abstract}
\noindent The global chaos caused by the 19 July 2024 technology meltdown highlights the need for a theory of what large-scale cohesive behaviors -- dangerous or desirable -- could suddenly emerge from future systems of interacting humans, machinery and software including AI; when will they   emerge; and how will they evolve and be controlled. Here we offer answers by introducing an aggregation model that accounts for the interacting entities' inter- and intra-species diversities. It yields a novel multi-dimensional generalization of existing aggregation physics. We derive   exact analytic solutions for the time-to-cohesion and growth-of-cohesion for two species, and some generalizations for an arbitrary number of species. These solutions reproduce -- and offer a microscopic explanation for -- an anomalous nonlinear growth feature observed in various current real-world systems. Our theory suggests good and bad `surprises'  will appear sooner and more strongly 
as humans-machinery-AI etc. interact more -- but it also offers a rigorous approach for understanding and controlling this. 
\end{abstract}

\maketitle

Systems containing different types of interacting humans, machinery (e.g. sensors,  actuators) and software including AI, look set to feature across societal domains ranging from commerce, medicine and defense to space exploration \cite{medicine,NatAcad,CSET,NAS,DOD,NSF,Jamie,Nancy,meta}.
Already, Metaverse-ready social media platforms allow individuals from anywhere in the world to spontaneously aggregate into a cohesive unit (social group \cite{meta,Palla,Artime,signal}) to perform activities (e.g. games, missions, economic transactions, goods exchanges) across multiple servers assisted by physical devices (e.g. headset \cite{Quest}) and interactive software/AI \cite{meta}. 
The interacting entities for the AI/machine-learning portion include
different categories of data tokens or topics extracted from unstructured multimodal datasets of text, images, audio, video  \cite{Rick,RickRick,Nanda1,Nanda2,Nanda3,acharya2022gensyn,Shehu,Axtell,jp,jay,danny}. Low-latency communications networks mean the interactions between such diverse entities will be largely independent of physical separation \cite{meta,Palla}. 

This raises key questions for both Society and fundamental science around what large-scale cohesive behaviors may suddenly appear from out-of-nowhere in such multi-species systems? When?  How will they evolve and be controlled? And what novel insights can Physics
add to address these questions? The spontaneous, bottom-up emergence of system-level cohesion may be desirable, e.g. emergence of cooperation and coordination in a human-AI-robot surgical team or space mission \cite{medicine,CSET}. But it may be undesirable and even highly dangerous, particularly if such systems belong to adversaries, e.g. emergence of AI-assisted online extremism \cite{BergerPerez,MillerIdriss,JohnsonScience2016}. Our use of the term `cohesion' here is consistent with the huge body of existing physics on polymers, networks and also social systems \cite{Jusup,Artime,signal,Palla,Stockmayer43,Stockmayer44,Flory1953,Ziff1982,Hendriks1983,Wattis2006,SMT1, Lushnikov06, Nelson2020,krapivsky,barabasi_book,newman,HalpinHealy,Schweitzer,Perra,Dodds,Mackay,Battiston21}, i.e.  sudden, spontaneous emergence and growth of some macroscopic unit (`cohesive unit') containing a non-negligible fraction of the entire multi-species population. In more traditional physics settings with identical entities, this `cohesive unit' represents the well-known gel, or giant connected component (GCC), or it could be a cartoon of some dynamically emerging coherent many-body state \cite{Halperin,krapivsky,barabasi_book,newman,HalpinHealy,Schweitzer,Perra,Dodds,Mackay,Erdos1959,Erdos1960}. 

\begin{figure}[t]
    \centering
    \includegraphics[width=0.75\linewidth]{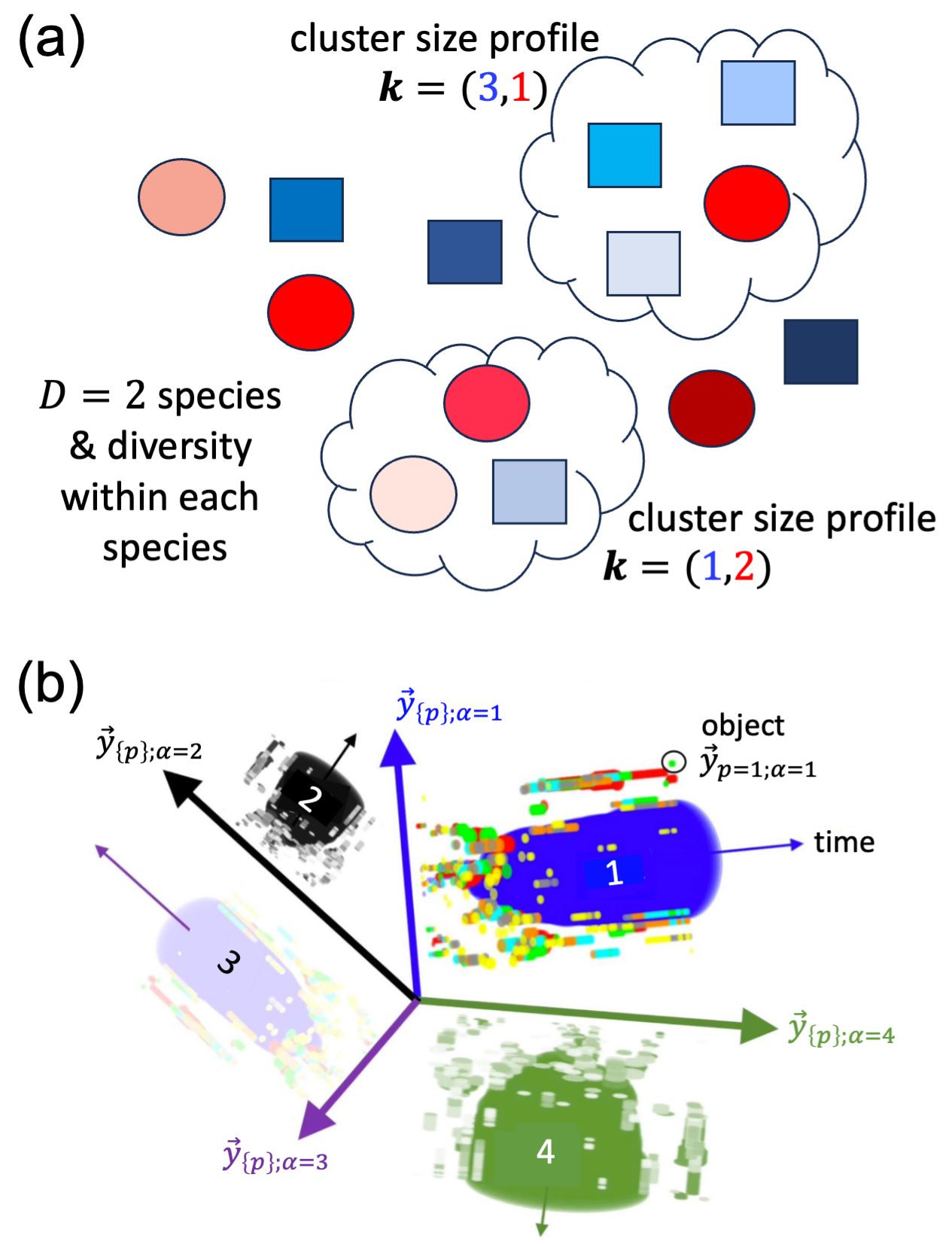}
    \caption{(a) Schematic of our theory at a given timestep for $N=13$ interacting entities from $D=2$ species (circle, square). Spectrum of given color represents intra-species traits for entities of that species. 
    (b) Our theory's output in absence of cross-species aggregation (i.e. off-diagonal $\mathsf{F}$ elements all zero). Adding cross-species aggregation (i.e. some or all $\mathsf{F}$ elements non-zero), our theory predicts $1,2,\dots D$ cohesive units emerge along different (mixed) directions, each with its own onset time and mixed-species composition.}
    \label{fig:2}
\end{figure}

Here we offer answers to these questions by introducing a multi-species physics model that permits some exact analytic solutions and reproduces the common feature observed across various current systems of a sudden rise and subsequent growth with $0,1,2,\dots$ kinks. 

The novelty of our theory lies in allowing for any number $D$ different species of entity, as well as diversity of the individual entities within each species (Fig. 1(a)).
The term `species' represents some major difference between entities, e.g. different types of humans, machinery, AI.
Hence it expands formal physics understanding by showing how existing dynamical equations for interacting identical particles/nodes \cite{Stockmayer43,Stockmayer44,Flory1953,Ziff1982,Hendriks1983,Wattis2006,SMT1, Lushnikov06, Nelson2020,krapivsky,barabasi_book,newman} generalize to new application domains (e.g. human-machinery-AI) involving $D>1$ species and within-species diversity. Numerical simulations confirm that our exact and approximate analytic results in this paper are accurate (SM Fig. 7).

Our theory starts with a general system containing $N=\sum_{\alpha=1}^D N_{\alpha}$ indivisible entities of $D$ major types (species) where the species label $\alpha$  denotes a dominant characteristic of each entity that does not change significantly over time. Different species could represent different types of humans or machines, or the same type of machines (software/AI) from different manufacturers (developers); or different categories of tokens or topics within AI/Large Language Models like GPT, DALL-E. Each entity $p$ within a given species $\alpha$ can have an arbitrary number of additional intra-species traits which could in principle change over time, denoted as $\vec{y}_{p;\alpha}(t)$. Each component (trait value) lies between 0 and 1 \cite{PhysRevLett.130.237401,
HalpinHealy,Gavrilets,PNAS,Centola,Char1,Char2,Char3,prl18}.

Similar to embedding in machine learning and AI, each entity is hence a vector in some high dimensional space. Some dimensions denote major characteristic differences (i.e.  species label $\alpha$); others represent the diversity within each species (Fig. \ref{fig:2}). 
As a result of interactions, some entities may find themselves at any time either coordinated, correlated or just interrelated in some way, hence forming effective clusters that can change in size, composition and total number over time (Fig.\ \ref{fig:2}(a)). The {\em cluster size profile} for each cluster at a given timestep is $\vb*{k} = (k_1, k_2,\dots k_D)$ where $k_\alpha$ is the number of species-$\alpha$ entities. Each cluster's total size is $k = \sum_{\alpha=1}^D k_\alpha$.
The number of clusters with the same size profile $\vb*{k}$ is $n_{\vb*{k}}$. Since $N_{\alpha} \gg 1$, we take $N_\alpha = N/D$ for simplicity. If all cross-species interactions are zero, we would see up to $D$ separate cohesive units (hence up to $D$ good or bad `surprises') emerge spontaneously over time, each of which contains a single species of entity (Fig.\ \ref{fig:2}(b)).

References \cite{PhysRevLett.130.237401, Palla, Erdos1959, Erdos1960} showed that the empirical aggregation mechanism in various human/machine(bots)/online-communications systems has an approximate product kernel form which we therefore adopt, though this can be easily generalized (see SM Sec. VIIB). Specifically, 
two randomly chosen entities fuse together at each timestep -- and if they are already part of a cluster, their entire clusters fuse together. We add here the additional feature that the fusion probability is some general function of the pair's similarity $S_{\alpha\beta}(\vec{y}_{1;\alpha},\vec{y}_{2;\beta},t)$ (e.g. \ $|\vec{y}_{1;\alpha}(t)-\vec{y}_{2;\beta}(t)|$) since cross-species and intra-species aggregation will generally occur at different rates.   
A $D$-dimensional fusion probability matrix $\mathsf{F}$ can be calculated by averaging the pair similarity function $S_{\alpha\beta}(\vec{y}_{1;\alpha},\vec{y}_{2;\beta},t)$ over the population distribution (SM Sec. IIA). Even if some $\vec{y}_{p;\alpha}$ terms vary quickly in time, the population average quantities in 
$\mathsf{F}$ will likely not. For non-binary fusion (i.e. multi-body fusion from coordinated aggregation such as influence campaigns), $\mathsf{F}$ will be a higher-order tensor. 

The general form for the cluster fusion dynamics within this mixed-entity (e.g. humans, machinery, AI) system then becomes (SM Sec. IIA):
\begin{equation}
    \dv{{\nu}_{\vb*{k}}}{\tau} = \frac{1}{2D}\sum_{\vb*{k} = \vb*{i} + \vb*{j}} 
            \vb*{i}^\mathrm{T}\mathsf{F}\vb*{j} \nu_{\vb*{i}}\nu_{\vb*{j}} - \vb*{k}^\mathrm{T}\mathsf{F} \left( \frac{\vb*{N}}{N} \right) \nu_{\vb*{k}} + g(\vb*{k},\tau)
    \label{eq:Dd_main} 
\end{equation}
where the dimensionless quantity $\nu_{\vb*{k}} = (D/N)n_{\vb*{k}}$ and $\tau=2t/N$ is a rescaled time. A function $g(\vb*{k},\tau)$ appears if higher-order interactions are included, e.g. $\geqslant 3$-body aggregation processes such as in coordinated influence campaigns (SM Sec. VIIB). As well as generalizing the Smoluchowski equation \cite{Wattis2006}, Eq.\ (\ref{eq:Dd_main}) is exactly equivalent to a $D$-dimensional generalization of the inviscid Burgers' equation plus additional shear interactions (SM Sec. IIB using  $\varepsilon_\alpha(\vb*{x},\tau) = \sum_{\vb*{k}}k_\alpha\nu_{\vb*{k}}e^{-\vb*{k}\cdot\vb*{x}}$): 
\begin{equation} \label{eq:generative_main}
    \pdv{\varepsilon_\alpha}{\tau} + \frac{1}{D}F_{\beta\gamma}(\varepsilon_\beta - \iota_\beta)\partial_\gamma\varepsilon_\alpha - \partial_\alpha \Tilde{g}(\vb*{x},\tau) = 0\ \ .
\end{equation}
 
\noindent The constant vector $\iota_\beta = 1\ \forall \beta = 1,\dots,D$.  $\Tilde{g}(\vb*{x},\tau)$ is the Laplace transformation of $g(\vb*{k},\tau)$. 

\begin{figure}[t]
    \centering
    \includegraphics[width=0.75\linewidth]{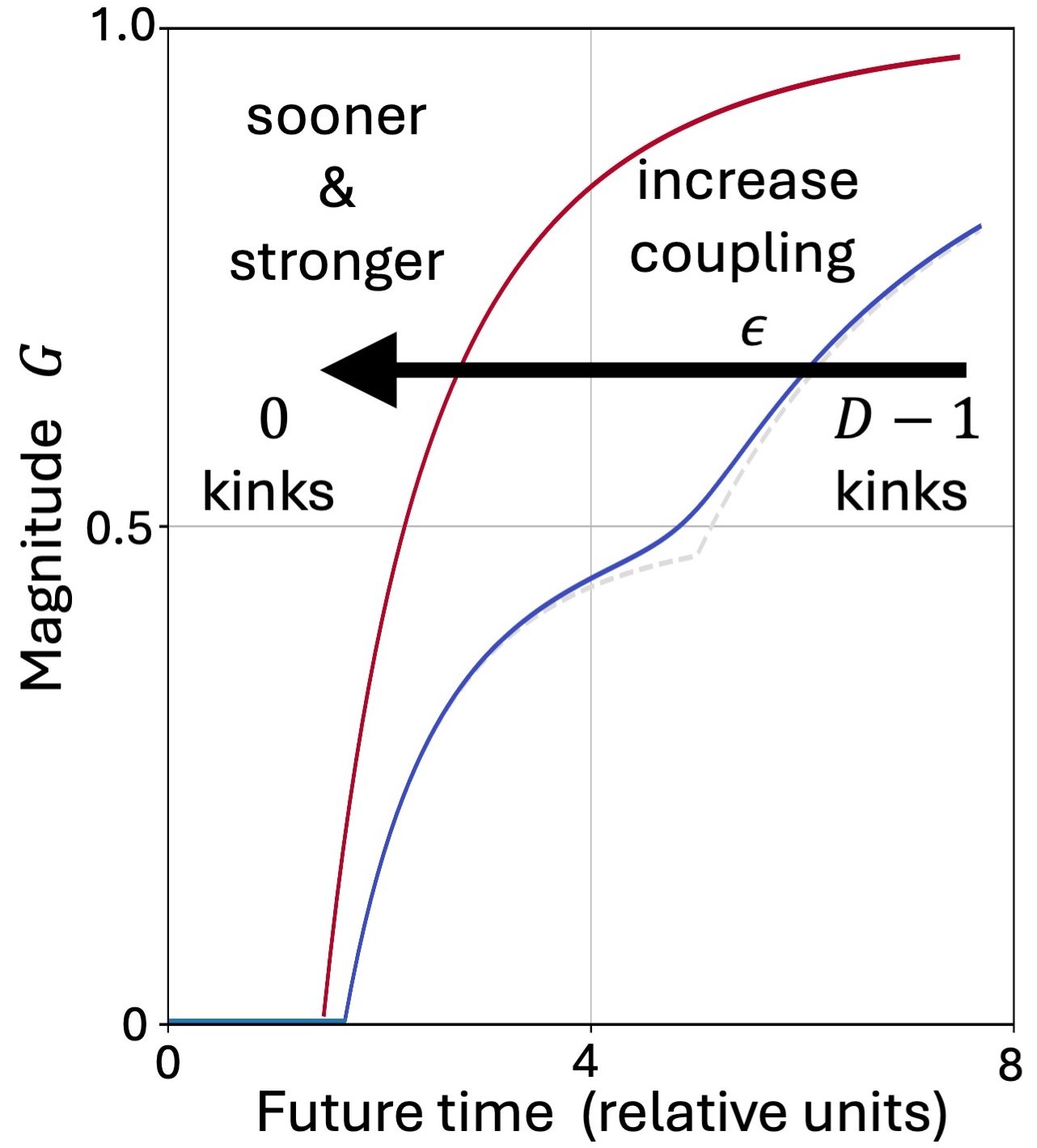}
    \caption{Our theory's predicted cohesive unit magnitude $G$ versus (scaled) time for $D=2$ species system ($G = (1/2) \sum_{\alpha=1}^2 G_\alpha(\tau)$). Cross-species aggregation term (i.e. off-diagonal terms in $2\times 2$ matrix $\mathsf{F}$) all set to $\epsilon$. As $\epsilon$ increases from 0 (gray dashed line)  the physically observed onset moves to earlier times, and the number of kinks reduces successively from $D-1=1$ to $0$. This means the rise in $G$ -- and hence a good or bad `surprise' -- will occur sooner and more strongly.}
    \label{fig:3}
\end{figure}

The solution to Eq.\ (\ref{eq:generative_main}) at $\vb*{x}=0$ gives the time-dependent growth of the mixed-species cohesive unit(s) that will emerge: $G(\tau) = 1-\sum_{\alpha=1}^D \varepsilon_\alpha(\vb*{0},\tau)/D$. For binary fusion (i.e. $g(\vb*{k},\tau) = 0$) and all-monomer initial condition, we solve Eq.\ (\ref{eq:generative_main}) analytically (SM Sec. IIB):
\begin{equation} \label{eq:gel_main}
    G(\tau) = 1 - \frac{1}{D}\sum_{\alpha=1}^D e^{-\sum_\beta F_{\alpha\beta} G_\beta(\tau)\tau/D}
\end{equation}
\noindent 
where $G_\beta(\tau) \equiv 1-\varepsilon_\beta(0,\tau)$ is the proportion of species-$\beta$ entities in the cohesive unit. 
A given cohesive unit $\mu$'s onset time (i.e. time-to-cohesion) is given by the eigenvalue $\lambda_\mu$ of $\mathsf{F}$ where $\mu=1,\dots,D$, i.e., $(\tau_g)_\mu = D/\lambda_\mu$. Its composition is given by the corresponding eigenvector $\vb*{x}^{(\mu)}$, i.e. 
$\vb*{k}^{(\mu)}$ such that $\vb*{k}^{(\mu)}\cdot\vb*{x}^{(\nu)} = \delta_{\mu\nu}$. 
$\vb*{k}^{(\mu)}$ represents a real-world cohesive unit only if all its components are non-negative. Up to $D$ such mixed-species cohesive units {\em could} emerge -- but any cohesive unit with an infinite or negative onset time will never appear.

This has a key real-world implication: we can predict the number of future cohesive units -- and hence good or bad `surprises' -- that will emerge, and their onset times, by analyzing the properties of $\mathsf{F}$. (Their growth can then be calculated using Eq.\ (\ref{eq:gel_main}) or more generally Eq.\ (\ref{eq:generative_main})).  The process proceeds by first
organizing $\mathsf{F}$ into block-diagonal form. Each submatrix $\mathsf{F}_s$ ($s=1,\dots,d$ where $d$ is the number of submatrices) then represents a separate  and generally mixed-species cohesive unit. In the absence of cross-species aggregation (i.e.\ diagonal $\mathsf{F}$ as in Fig. \ref{fig:2}(b)) $d=D$ and each cohesive unit exhibits the well-known single species growth curve.  
For cases where each $\mathsf{F}_s$ is symmetric and non-negative, the Perron-Frobenius theorem ensures that it has an only has one non-negative eigenvector, labelled as $\vb*{x}_\textnormal{max}^{(s)}$, corresponding to the largest eigenvector $\lambda_\textnormal{max}^{(s)}$. 
Hence for each irreducible subsystem $\mathsf{F}_s$, there exists exactly one onset of a cohesive unit at $\tau_g^{(s)} = D/\lambda_\textnormal{max}^{(s)}$, with composition given by the $\vb*{k}_\textnormal{max}^{(s)}$. Each can be found exactly. 
This can be understood by imagining gradually turning on all the cross-species aggregation entries starting from a diagonal $\mathsf{F}$ (i.e., off-diagonal entries become non-zero): hence $\mathsf{F}$ finally becomes irreducible which mathematically has the effect of rotating the principal axes of $\mathsf{F}$ until all but one of the eigenvectors have at least one negative component -- hence only one mixed-species cohesive unit would emerge. The total number of emerging cohesive units (future `surprises') will hence lie between $1$ and $D$.

 An exact analytical solution can be derived for any number of species $D$ in the case that all intra-species terms are identical (i.e. all diagonal entries in $\mathsf{F}$ equal to $f$) and all inter-species terms are identical (i.e. all off-diagonal entries equal to $\epsilon$). The Perron-Frobenius theorem ensures that $\vb*{k}^{(0)} = (1,\dots,1)/\sqrt{D}$ is the only cohesive unit that will emerge, with an exact onset time (derived in SM Sec. VI):
\begin{equation} \label{eq:effective_main}
    \tau_g = D[f+(D-1)\epsilon]^{-1} 
\end{equation}
which also acts as a good approximation when the relevant $\mathsf{F}$ entries are similar have average values $f$ and $\epsilon$. 

Key practical implications of Eq.\ (4) are: 

\noindent (i) cohesive units (and hence good and bad `surprises') will appear {\em sooner}  
as humans-machinery-AI etc.\ mix more (i.e.\ $\tau_g$ decreases as $\epsilon$ and/or $f$ increase), and they will also appear {\em stronger} since the various contributions pile up as shown in Fig. 2 for $D=2$; 
(ii) as the number of species increases ($D\rightarrow\infty$), $\tau_g\rightarrow \epsilon^{-1}$ and hence the system eventually behaves like a renormalized one-species system in which the intra-species aggregation probability is now $\epsilon$ instead of $f$. $f$ becomes irrelevant. 

\begin{figure}[h]
    \centering
    \includegraphics[width=0.8\linewidth]{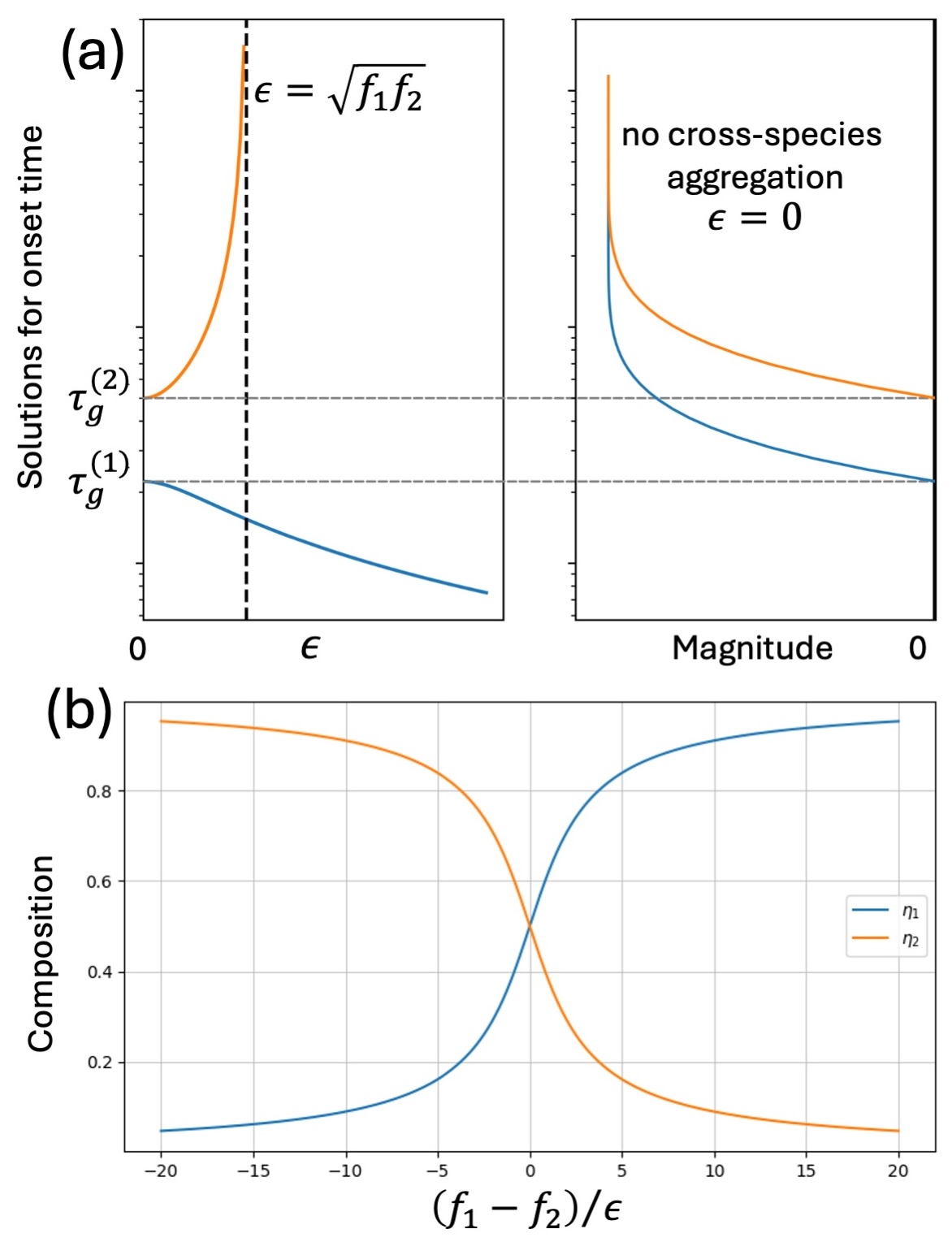}
    \caption{(a) Left: Solutions for onset time(s) as a function of inter-species interaction $\epsilon$ for $D=2$ species. Onset time of overall growth curve $G$ is the smallest (i.e. earliest) solution $\tau_g^{(1)}$. Right plot depicts the separate contributions to the overall cohesive unit $G = (1/2) \sum_{\alpha=1}^2 G_\alpha(\tau)$ at $\epsilon=0$. 
    (b) Proportions of the two species as a function of $\sigma \equiv (f_1-f_2)/\epsilon$: larger $\sigma$ values denote increasing relative asymmetry in the intra-species aggregation probabilities and hence higher asymmetry in the cohesive unit's composition. SM Secs. IV, V give all details of analytic solutions for $D=2,3$.}
    \label{fig:2d_time_comp}
\end{figure}

Figure 2 shows the simple two-species ($D=2$) solution from Eq.\ (\ref{eq:gel_main}): it predicts the appearance of a kink which then gets smeared out as $\epsilon$ increases. See SM Secs. IV, V for details of $D=2,3$ analytic solutions and SM Fig. 5 for $D=3$ case. This provides an endogenous explanation of why a kink  appears: a subset of entities from across species get involved immediately while the others act collectively {\em as if} they are sitting on-the-fence, even though there is no such individual-level delay mechanism in the model. Instead, the delay is a collective phenomenon. More generally, the $D$-dimensional structure of $\mathsf{F}$ means that a system of $D$ species will produce $n_G=1,2,\dots,D$ separate growth curve solutions, where each can have $0,1,2,\dots,D-n_G$ kinks. This suggests that a system exhibiting growth curves with $0,1,2,\dots D-n_G$ kinks, is effectively behaving as a system of $n_G$ renormalized species. Each renormalized species might correspond to some seemingly obvious category (e.g. all humans) but more generally it will be a complex subset of entities (e.g. some humans and some sensors/actuators) that would have been hard to identify a priori.

Consider $D=2$ with {\em non}-identical diagonal $\mathsf{F}$ elements $f_1,f_2$. 
The $2\times 2$ $\mathsf{F}$ matrix has two roots (SM Sec. IV.A): 
{\small\begin{equation} \label{eq:gel_time_2D}
    \tau_g^{(1,2)} = [{(f_1+f_2) \mp \sqrt{(f_1-f_2)^2 + 4\epsilon^2}}][{f_1f_2-\epsilon^2}]^{-1}\ \ .
\end{equation}}

\noindent Figure 3(a) shows that the variation of these two roots with  $\epsilon$ is 
strikingly similar to the hybridization in a 2-band model from solid state physics \cite{kdotp}.  
When $\epsilon=0$, the two species form separate single-species cohesive units at times $\tau_g^{(1,2)} = 2/f_{1,2}$. 
Turning on the cross-species interaction (i.e. $\epsilon>0$) there is a transition such that only one mixed-species cohesive unit emerges at $\tau_g^{(1)}$. 
As $\epsilon$ grows larger, $\tau_g^{(1)}$ becomes smaller hence the mixed-species cohesive unit forms earlier in agreement with Fig. \ref{fig:3}. By contrast, the larger root $\tau_g^{(2)}$ increases: but instead of a real onset, it denotes a `virtual' one driven by the latency of the slower-aggregating species -- hence the kink.
$\tau_g^{(2)}$ then diverges at $\epsilon^2=f_1f_2$. For $\epsilon^2>f_1f_2$ there is only one positive root and the kink disappears. 
An analytic solution for the cohesive unit's composition just after onset, comes from either Taylor expanding Eq.\ (\ref{eq:gel_main}) or calculating $\vb*{k}^{(1,2)}$ (see SM Sec. IV.B). The composition of species $\alpha$ in this cohesive unit $\eta_\alpha$ (see Fig.\ \ref{fig:2d_time_comp}(b)) depends solely on $\sigma \equiv (f_1-f_2)/\epsilon$:
    \begin{align}
        \eta_1 &= {2}\left[{\sqrt{\sigma^2+4}-\sigma+2}\right]^{-1} \\
        \eta_2 &= 1-\eta_1 = {2}\left[{\sqrt{\sigma^2+4}+\sigma+2}\right]^{-1}\ \ .
    \end{align}
   The cohesive unit's composition is hence highly sensitive to small asymmetries in the two intra-species aggregations (i.e. $\sigma$). This has a key real-world implication: too much symmetry in different species' internal aggregation probabilities will make the cohesive unit's composition unstable, i.e. it becomes risky to assume that it has a particular quasi-static composition.
    
Large-scale human-machine-software/AI systems are yet to be developed, hence the best application scenarios for our model do not yet exist. However, Fig. 4 shows that even the simplest (i.e. $D=2$) version of our theory reproduces the growth curves and onset times in various current candidate systems. 
This good agreement in Fig. 4 is highly non-trivial. If the onsets and growth kinks were caused by exogenous news event(s) in panels (a) and (d), they should be synchronous -- but they are not, so an exogenous cause is unlikely. Also the real-world setups in (b)-(d) make exogenous causes practically impossible.
 Our theory of multi-species heterogeneity hence yields deterministic predictions of multi-kink growth profiles and onset times without having to search for exogenous causes, and it sits well beyond existing theories that feature percolation among identical objects. Not all kink-like growth curves will always be completely endogenous and hence candidates for our theory, but we expect a sizeable number of future systems will be.
 
\begin{figure}[h]
\includegraphics[width=0.92\linewidth] {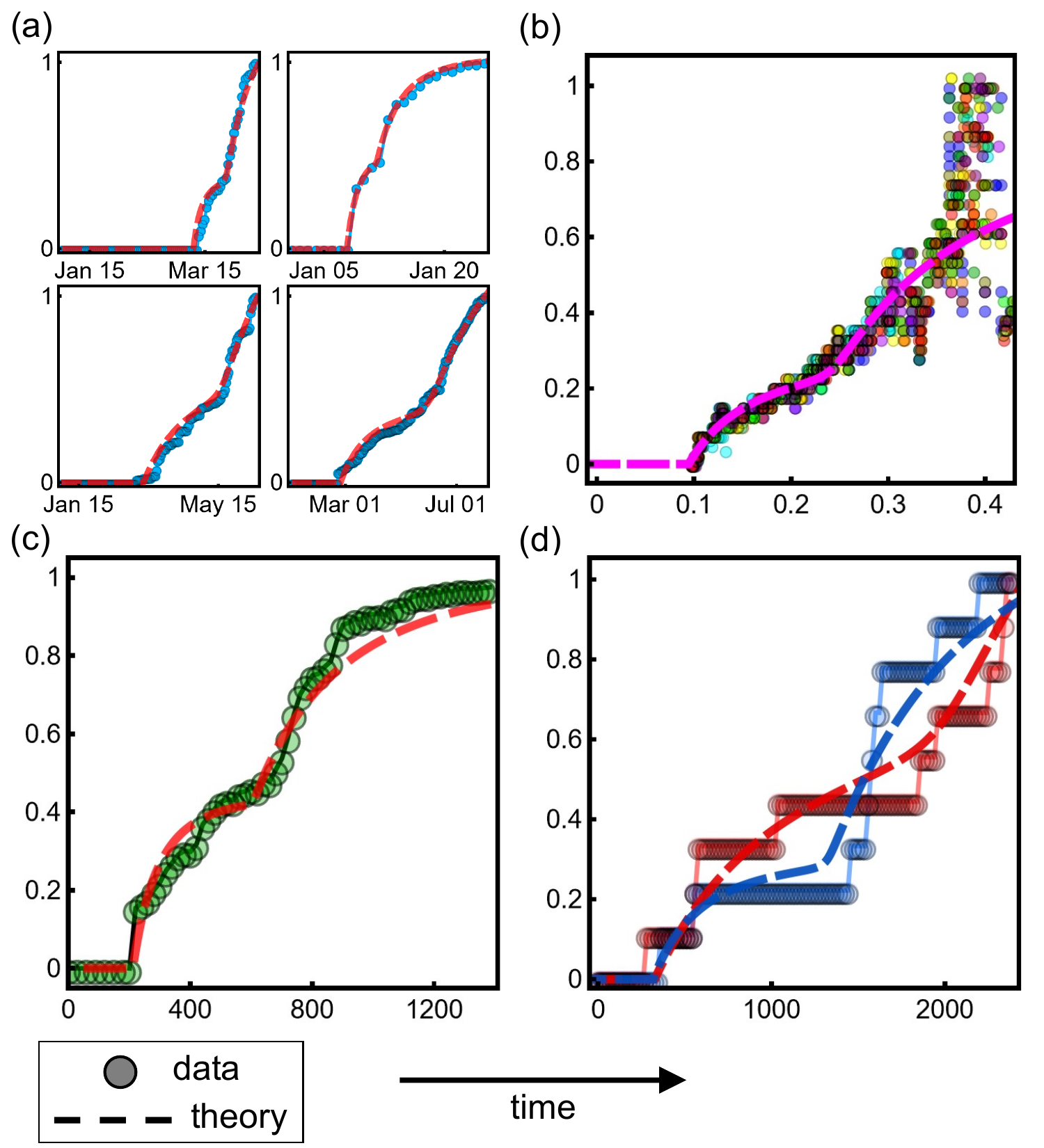}
    \caption{Sudden rise and multi-kink growth of a macroscopic cohesive unit with magnitude $G$ (vertical axis) vs. time (horizontal axis) in various current real-world systems of interacting humans, machinery, software compared to the simplest version of our theory (i.e. $D=2$). See End Matter for details of data, normalization and theory curve fits: (a) online anti-U.S. jihadi group membership during given period \cite{JohnsonScience2016}; (b) subsecond human-machine trading across electronic exchanges \cite{NeilScience}; (c) reduction in loss function during AI learning \cite{Nanda1,Nanda2,Nanda3}; (d) human-machine-software activity during two nominally identical pilot space missions \cite{Jamie}.}
\label{fig:motivate}
\end{figure} 

Beyond the Physics, this work offers a takeaway for Society: good and bad `surprises' (i.e.,\ sudden emergence of cohesive units) will likely appear sooner and stronger 
as humans-machinery-AI etc. mix more, and a Physics approach such as this can offer a rigorous understanding and test-bed for control interventions.  
Our multi-species theory is obviously simplistic: however this makes it generic enough to apply at many scales and with many types of entities. We also speculate that its focus on entities' (tokens') assembly could help provide a first microscopic explanation for the rather abrupt appearance of output abilities in machine learning, AI and new Transformer architectures (see Refs.\cite{Nanda1,Nanda2,Nanda3}) due to sudden cohesion among the abstract circuits that develop within the machine-software system. The caveat is that the sequence of cluster assembly must then be included using more complex temporal terms in $\mathsf{F}$. 

This research is supported by U.S. Air Force Office of Scientific Research awards FA9550-20-1-0382 and FA9550-20-1-0383 and The Templeton Foundation.




\bibliography{2_species_agg}

\onecolumngrid

\vskip1in 
\begin{center}
    {\large \textbf{End Matter}}
\end{center}

\twocolumngrid
 
In Fig. 4, the vertical axis gives the normalized size $G$ of the coherent unit that emerges in each real-world example. (a) Online anti-U.S. jihadi group membership during a given period. The system involves members and potential recruits interacting with the online platform VKontakte's servers, algorithms and software which run the system \cite{JohnsonScience2016}. These realizations are from the same time period. 
(b) Sub-second human-machine trading across electronic exchanges. Horizontal scale is in seconds. Each color is the price of a given stock on a different electronic exchange \cite{NeilScience} during the same time period. The price measures the aggregate action of different types of human traders and bot traders (i.e., software/algorithmic). 
(c) Reduction in loss function during AI learning associated with generalization, i.e. grokking \cite{Nanda1,Nanda2,Nanda3}. Horizontal scale is iterations over time. The model producing the data is a Transformer trained on the problem of 5 digit addition \cite{Nanda1,Nanda2,Nanda3}. The loss function $L(I)$ decreases with the number of iterations $I$. We take the initial value $L(0)$ and construct the metric $(L(0)-L(I))/L(0)$ which is proportional to $G$ during the growth period, i.e., $t>200$. (d) Human-machine-software activity during two nominally identical pilot space missions \cite{Jamie}. Horizontal scale is in seconds. Each mission's progress is measured by the number of accumulated sensor events up to that instant across all its human-machine-software sensors (i.e. $G$).
The normalizations for panels (b) and (d) in Fig. 4 are done as follows. For (b): let $\bf X$ be the list of prices as a function of time and $x_j$ is its $j$'th entry. Each new (i.e. normalized) entry $x_{j'} = (x_{j} - min({\bf X}))/(max({\bf X})-min({\bf X}))$ where the max and min are global across all lists. For (d), we simply divide each entry by the maximum for both cases. Full details of these data are given in SM Sec. I.

Because our intention in Fig. 4 is simply to show that our theory is capable of reproducing the kink-like growth seen empirically, the theory curves in Fig. 4 are obtained from the simplest form of our multi-species theory (i.e., $D=2$) which gives one or zero kinks. The SM Sec. IV gives explicit expressions for these growth curves. Choosing $D\geq 3$ would allow better fits to capture additional minor kinks, such as those seen in Fig. 4(c). For each panel, we proceed to plot our $D=2$ theory results as follows. First, we try an initial exploratory fit to the empirical data with our multi-species theory, specifically the general Eq.\ (41) in the SM for $D=2$ using an estimate of the number of objects based on the specifics of the data, e.g., the online group membership. This initial comparison then allows us to estimate a suitable time transformation between a timestep in the theory and real time as in the data: e.g., for the top left panel of Fig.\ 1(a) this yields $600$ timesteps $\approx$ 1 day. We then find the entries of the $\mathsf{F}$ matrix, as well as the population sizes for each species, that together give the best match to the kink positions. The growth curves themselves are dictated solely by the theory (Eq. (41)). The abruptness of the kinks indicates the strength of the inter-species interactions compared to the intra-species interactions. 
While it is possible that other fitting procedures provide a better fitting statistic, they will not necessarily capture the kink structure which is our main focus. Our use of the minimal $D=2$ case also helps us avoid over-fitting.

\end{document}